\documentclass[aps,prb,twocolumn,superscriptaddress,10pt]{revtex4}

\newcommand{\comment}[1]{}

\usepackage{amsmath}
\usepackage{amssymb}
\usepackage{latexsym}
\usepackage{indentfirst}
\usepackage{graphicx}
\usepackage{color}
\usepackage{bm}
\usepackage{multirow}

\begin{document}

\title{Audio Cards for High-Resolution and Economical Electronic Transport Studies}

\author{D.~B. Gopman}
\author{D. Bedau}
\email{db137@nyu.edu}
\author{A.~D. Kent}
\affiliation{Department of Physics, New York University, New
             York, NY 10003, USA}

\begin{abstract}
We report on a technique for determining electronic transport properties using commercially available audio cards. Using a typical 24-bit audio card simultaneously as a sine wave generator and a narrow bandwidth ac voltmeter, we show the spectral purity of the analog-to-digital and digital-to-analog conversion stages, including an effective number of bits greater than 16 and dynamic range better than 110 dB. We present two circuits for transport studies using audio cards: a basic circuit using the analog input to sense the voltage generated across a device due to the signal generated simultaneously by the analog output; and a digitally-compensated bridge to compensate for nonlinear behavior of low impedance devices. The basic circuit also functions as a high performance digital lock-in amplifier. We demonstrate the application of an audio card for studying the transport properties of spin-valve nanopillars, a two-terminal device that exhibits Giant Magnetoresistance (GMR) and whose nominal impedance can be switched between two levels by applied magnetic fields and by currents applied by the audio card. 
\end{abstract}


\maketitle

Studies of the electronic transport properties of devices and materials require specialized instrumentation. Most transport studies break down into two classes: current-voltage characterization and differential resistance measurements. The former can be done using a sourcemeter, which supplies power to a device and simultaneously measures both the voltage drop across the device and current flowing through the device. Differential resistance is typically measured using a lock-in amplifier. A lock-in amplifier is a robust solution for accurate, low-noise measurements of the voltage response $\delta U$ to a small excitation current, $\delta I$. This type of measurement has a narrow noise bandwidth and can be conducted above the 1/$f$ noise threshold with high dynamic reserve. However, it is sometimes desirable to measure the static response (current-voltage characteristics) simultaneously with the differential response, $\delta U$/$\delta I$. This cannot be done with only a lock-in amplifier and requires the addition of an additional source-measure instrument. Furthermore, commercially available scientific instruments for both measurement tasks can cost upwards of \$10,000. 

Commercially available audio cards can output and record arbitrary waverforms and do many of the tasks typically reserved for source-measure units and lock-in amplifiers. These are low-cost, versatile instruments, with 24-bit processing and functionality as a signal generator with less than -100~dBc harmonic distortion. Audio cards also can function as a narrow bandwidth ac voltmeter, with better than 110~dB dynamic range. With such high resolution, these devices outpace many direct digital synthesizers as a pure sine-wave generator and rank with commercial lock-in amplifiers in dynamic rejection. Furthermore, audio cards typically have multiple synchronized input-outputs which permit simultaneous characterization of multiple devices.

We present the application of an USB-based audio card - the E-MU 0404 (Creative Professional) - for static and dynamic transport measurements, which can be assembled very economically compared to commercial scientific instruments \cite{SoundCard}. This measurement device permits sourcing and detecting differential signals $\delta U$/$\delta I$ as well as sensing current-voltage characteristics simultaneously. We will present measurements of magnetic nanostructures, whose current-voltage and differential resistance are used to determine their underlying magnetic and electronic properties.

\begin{figure}[!b]
  \begin{center}
    \includegraphics[width=3.0in,
    keepaspectratio=True]
    {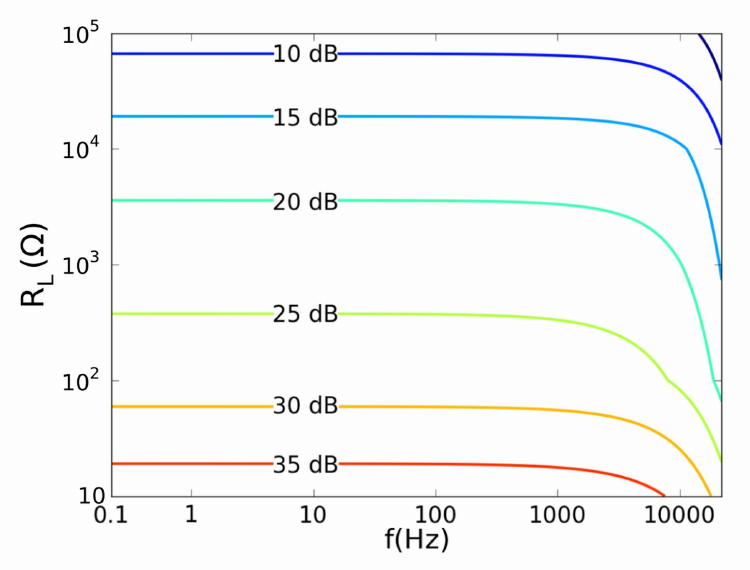}
  \end{center}
  \caption{\label{fig:nsd} Noise spectral density contours for load resistances up to 1~$M \Omega$ soldered directly to the audio card input. FFT spectra taken from the input data (bandwidth $\Delta f$ = 0.2~Hz) at each resistance is normalized by the corresponding Johnson-Nyquist noise, $\sqrt{4k_BTR_L}$, where $k_B$ is Boltzmann's constant, T is room temperature for these measurements, and $R_L$ is the load resistance.}
\end{figure}

Our measurement setup is centered around a single E-MU 0404 connected via USB to a PC. The audio card possesses two balanced high-impedance (1~M$\Omega$) inputs and two unbalanced analog outputs ($V_{pk}$ = 2 V). Analog-to-digital (A/D) and digital-to-analog (D/A) conversion is done with a variable sampling rate (44.1, 48, 88.2, 96, 176.4 and 192 kHz from an internal crystal) with 24-bit input/output (I/O) processing. Also included are optical digital interconnects (S/PDIF) for synchronous I/O over multiple audio cards. The data is read from the audio card using the Steinberg Audio Stream Input/Output (ASIO SDK) drivers, which are controlled using a Python wrapper (PyAudio) around PortAudio C programming libraries \cite{ASIO,PyAudio,PortAudio}.

\begin{figure}[t]
  \begin{center}
    \includegraphics[width=3.0in,
    keepaspectratio=True]
    {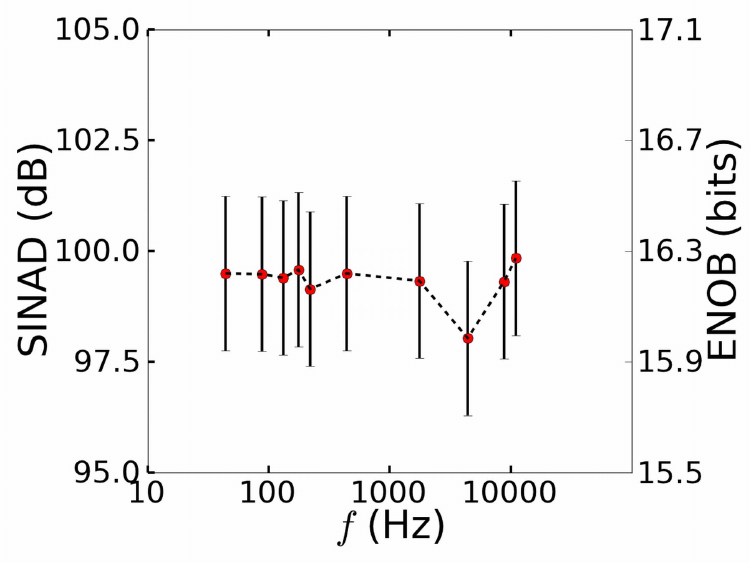}
  \end{center}
  \caption{\label{fig:sinad} Ratio of signal to noise and distortion (SINAD) and effective number of bits (ENOB) for signal frequencies ranging between 44.1~Hz and 11~kHz, demonstrating spectral purity of both input and output stages of the audio card. Output D/A channel is a sine wave that is looped back into one of the input A/D channels. FFTs are computed from recorded input signal.}
\end{figure}

We have tested some standard spectral characteristics of the audio card, many of which are described in electronics textbooks \cite{HORHILL}. Figure~\ref{fig:nsd} demonstrates the noise spectral density of the A/D converter at the input stage. For a variety of input resistances, the noise density generated from a fast fourier transform (FFT) of the input signal (bandwidth $\Delta f$ = 0.2~Hz) compared with the Johnson noise is nearly flat across the frequency spectrum (20~Hz-20~kHz). The noise amplitude gets closer to the Johnson noise for larger impedances, which could suggest that voltage noise at the input stage plays a role for the smaller load impedances. Furthermore, the input noise is only one order of magnitude larger than the best input noise for a commercial lock-in amplifier \cite{SR830}. Figure~\ref{fig:sinad} illustrates the spectral purity of the entire audio card by looping the output of the D/A converter (a sine wave at 6~dB below full scale) back into the A/D converter without preamplification. The calculated ratio of signal to noise and distortion (SINAD) from an FFT of the input signal leads directly to the effective number of bits (ENOB), which is a measure of the quality of the I/O digitization. A nearly constant value of 16 bits indicates that of the full 24 bits resolution, the upper 16 bits contain useful information above the noise floor, which is critical for sensing small signal changes over a larger background signal. Other useful spectral properties of the measured output include total harmonic distortion plus noise (THDPN) (-99.5 dBc (decibels below carrier)), signal-to-noise (-100.5 dBc), spurious free dynamic range (-112 dBc), and the noise floor (-132 dBc). Thus, we have a I/O device that functions as an ultra-low distortion, 24-bit resolution signal generator with an A/D input stage that measures low and higher-impedance devices alike with relatively flat noise density over the range of audible frequencies.

\begin{figure}[!b]
  \begin{center}
    \includegraphics[width=3.0in,
    keepaspectratio=True]
    {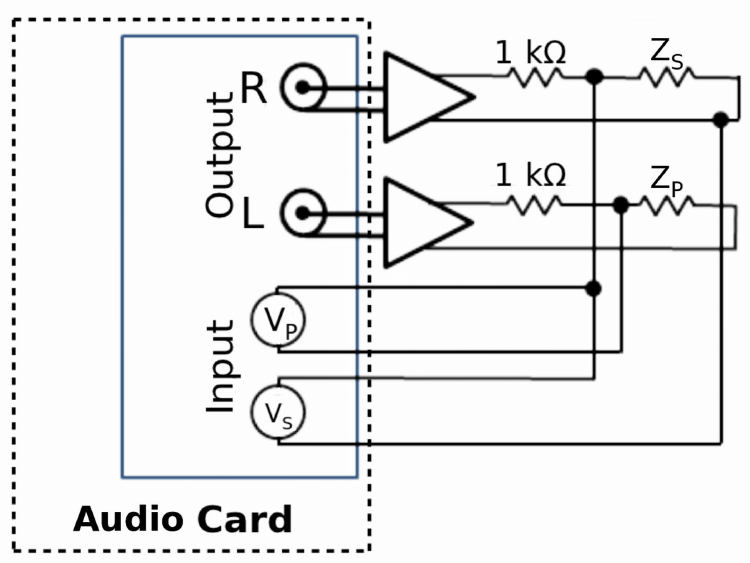}
  \end{center}
  \caption{\label{fig:schematic} Schematic of a digitally compensated bridge that uses two analog input and two analog output channels using the 24-bit processing I/O of a commercial USB audio card. $Z_S$ and $Z_P$ are the sample and potentiometer impedances of the bridge and are approximately two orders of magnitude lower in resistance than the 1~$k \Omega$ resistors. Outputs ``L'' and ``R'' are independent and can be adjusted along with $Z_P$ in order to compensate the signal from $Z_S$. Inputs ``$V_S$'' and ``$V_B$'' measure the voltage drop across the two arms of the bridge.}
\end{figure}

Figure~\ref{fig:schematic} illustrates the configuration we use for transport measurements. The top-left block represents the USB-based audio card, whose two outputs, L and R, are amplified and fed into two 1~$k \Omega$ resistors (chosen to be two orders of magnitude larger than the sample resistance) to create two stiff current sources, which feed into the sample ($Z_S$) and a potentiometer ($Z_P$). Leads across the terminals of $Z_S$ feed into the ``$V_S$'' audio card input and permit direct measurement of the sample I-V characteristic and/or differential resistance. 

We also implement a digitally compensated bridge measurement using the audio card. Balancing bridge designs based upon active compensation have been well established for systems where the versatility to compensate for arbitrary reactances in complex networks where capacitances and inductances need to be compensated, or even for systems where small nonlinear deviations in I-V characteristics require versatile and precise balancing \cite{jA66,dC72,mR85}. By outputing an appropriate signal on channel L, any parasitics or reactances of ``$Z_S$'' can be compensated for. This circuit permits us to compensate nonlinearities of the device I-V characteristic as well as unaccounted series reactive elements. It also functions as an inexpensive impedance meter across a wide frequency band (20 Hz - 20 kHz).  

\begin{figure}[!t]
  \begin{center}
    \includegraphics[width=3.0in,
    keepaspectratio=True]
    {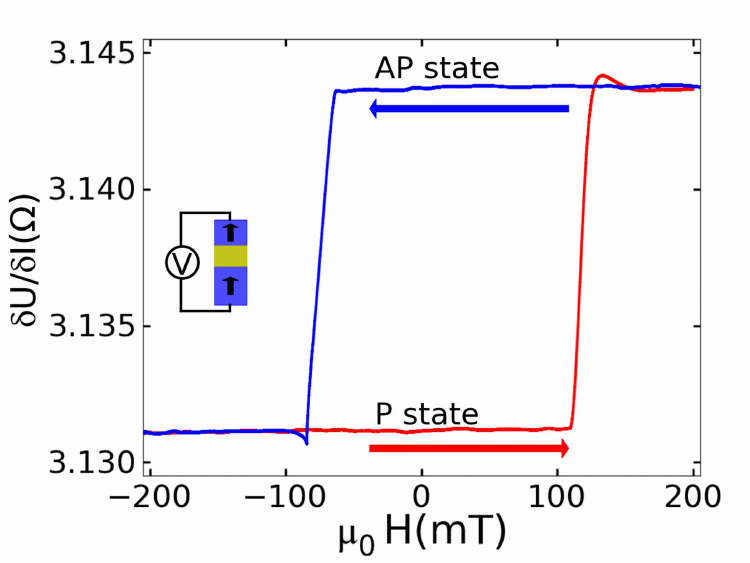}
  \end{center}
  \caption{\label{fig:RvsH} Lock-in amplifier measurement. A small ac excitation current (frequency = 882~Hz) is injected into a $300\times50~$nm$^2$ spin-valve nanopillar to probe $\delta U$/$\delta I$. The digitized voltage across the device is processed using a digital lock-in detection method. The differential resistance is plotted as a function of measured field. Sharp changes in $\delta U$/$\delta I$ indicate toggling between parallel and antiparallel magnetic orientations of the two ferromagnetic layers. (Inset) Schematic of a spin-valve nanopillar device attached to a voltage supply. Two ferromagnetic layers (blue) separated by a copper spacer (yellow) exhibit a high resistance state when their magnetizations are antiparallel and a low resistance state when their magnetizations are parallel.}
\end{figure}

Audio card measurements are especially useful for fast characterization of spin-valve nanopillars - a two terminal magnetic device composed of two uniaxial ferromagnetic layers that exhibits two stable resistance states depending on the relative magnetization orientation of the layers (see the inset of Figure~\ref{fig:RvsH}). The spin-valve state can be toggled between high (anti-parallel) and low resistance (parallel) by applied magnetic fields (field-induced magnetization switching) or by electrical currents (current-induced magnetization switching). We can determine the relative orientation of the layers by measuring the device resistance (static or differential) as a function of the applied magnetic field. However, the resistance change is small in comparison to the device resistance and the Joule heating-related nonlinearity, which requires that we measure small changes in resistance and differential resistance in order to observe the device switch between high and low resistance states.

Measuring the differential resistance versus field of a spin-valve nanopillar allows us to demonstrate the use of the audio card as a lock-in amplifier. We used the audio card to source a 882~Hz excitation current (200 $\mu$A rms amplitude) to our device, while simultaneously using the audio card to measure the generated voltage. We then multiplied the digitized voltage by a unit amplitude sine and cosine of the same frequency to get the in-phase and out-of-phase components of our signal in relation to our reference and use a digital, first-order Butterworth lowpass filter (time constant 30 ms) to eliminate noise and the $2 f$ component, and then calculate the magnitude from the in-phase and out-of-phase components, plotting $\delta U$/$\delta I$ as a function of applied magnetic field in Figure~\ref{fig:RvsH} \cite{Butter}. Sharp changes in the differential resistance indicate toggling between parallel and antiparallel magnetic orientations of the two ferromagnetic layers.

There is an interplay between applied magnetic fields and electric currents in the orientation of the two magnetic layers, which has been examined previously and is quite important for applications like magnetic random access memory \cite{db10a,sM06}. This relationship is usually plotted in the form of a state diagram, which maps out under a given applied magnetic field and direct current the boundary of different microscopic magnetization configurations of the two spin-valve layers as determined by sharp changes in measured giant magnetoresistance (GMR).

Thus far, studies of the spin-valve state diagram have been limited by the speed of the measurement, which can take several days to complete in order to explore the wide parameter space of applied magnetic fields and electric currents. We present a method for quickly mapping out the field-current parameter space of a spin-valve nanopillar with the use of the audio card in the direct measurement (I-V) configuration.

\begin{figure}[!b]
  \begin{center}
    \includegraphics[width=3.0in,
    keepaspectratio=True]
    {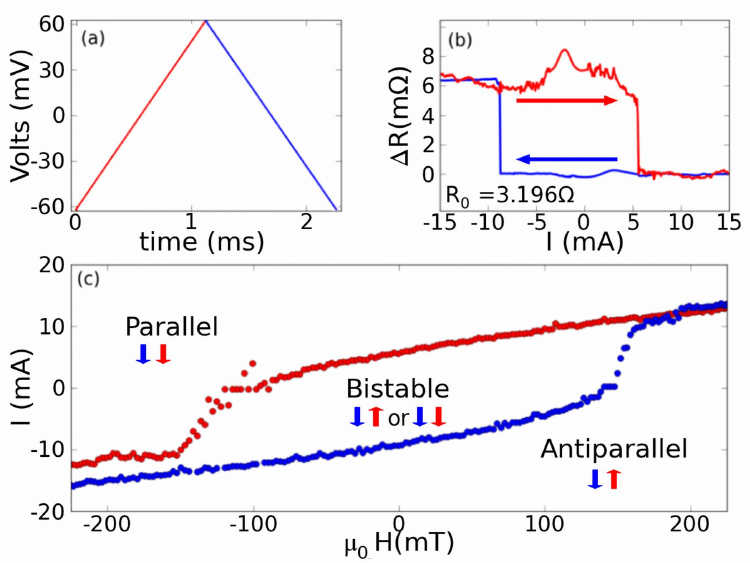}
  \end{center}
  \caption{\label{fig:magnetic} Magnetic measurements setup. (a) A 20~mA peak amplitude triangle wave flows through a spin-valve nanopillar device. The voltage measurement is conducted across the spin-valve terminals.  (b) Giant magnetoresistance hysteresis loop as a function of applied current showing sharp resistance changes between parallel (P) and antiparallel (AP) orientations of the two ferromagnetic layers. (c) (H,I) state diagram showing interplay between applied magnetic fields and electric currents in stable magnetic orientations of a spin-valve nanopillar.}
\end{figure}

In Figure~\ref{fig:magnetic}, our experimental results are displayed. We measure the voltage generated across the sample, $Z_S$, on channel ``$V_S$'', under a linear current ramp in the form of a 441 Hz triangle wave with peak ampllitude $I_{pk}$ = 20 mA flowing through a spin-valve nanopillar device whose cross sectional area is $300\times50~$nm$^2$. The physical properties of this device have been described in detail elsewhere \cite{db10b}. In Fig. \ref{fig:magnetic}(a) the voltage across the nanopillar is measured by the analog-to-digital converter (D/A) simultaneous to the application of the triangle wave from the audio card's digital-to-analog (D/A) converter). Figure~\ref{fig:magnetic}(b) demonstrates the corresponding GMR effect for current switching of the spin-valve from an antiparallel (high-resistance) to parallel (low-resistance) configuration and back again.

In order to map out the state diagram, we generate a triangle field ramp from our electromagnet with period T = 8 s and $H_{pp}$ = 200 mT on top of our continuous ac current. Over one half-period of the field cycle we can generate several hundred GMR curves like the one shown in Fig. \ref{fig:magnetic}(b), and by monitoring the currents and fields at which the device resistance toggles between high and low, we arrive at the state diagram in Fig. \ref{fig:magnetic}(c). This diagram illustrates the outer regions in which the spin-valve device is always oriented P or AP, and where the device exhibits bistability or hysteresis in the inner regions of the parameter (H,I) space.

In summary, we have demonstrated the use of an inexpensive, commercially-available audio card for high quality measurements of nanostructures. We believe that this can save researchers a significant amount of money on otherwise expensive standard scientific measurement tools while affording greater flexibility for signal processing, multiple channels for parallel studies, high dynamic range and most importantly, high speed capabilities. We have shown the spectral properties that make the E-MU 0404 USB Audio Card competitive with commercial laboratory instruments. The audio card unit is versatile - it can function as a tool for direct measurement of devices or it can be integrated into a digitally compensating bridge. While we have demonstrated the use of the audio card as a source-measure unit and as a digital lock-in amplifier for the study of low-impedance spin-valve devices, it should be noted that this measurement technique can be used for many types of measurements where a high quality signal source is required.

\section*{Acknowledgments}
We appreciate Dr. St\'{e}phane Mangin of Nancy Universit\'{e} and Dr. Eric E. Fullerton of the University of California, San Diego for providing the spin-valve samples used in the characterization of magnetic properties used in this study. We also acknowledge Marius Klein for support in the initial stages of developing these measurements. This research was supported at NYU by NSF Grant DMR-1006575. 

\bibliographystyle{apsrev}

\end{document}